\documentstyle[aps,floats,prb,epsfig]{revtex}

\begin{document}



\twocolumn[\hsize\textwidth\columnwidth\hsize\csname @twocolumnfalse\endcsname

\title{\bf Precise Tight-binding Description of the Band Structure
of MgB$_2$}

\author{D. A. Papaconstantopoulos and M. J. Mehl}

\address{Center for Computational Materials Science, Naval
Research Laboratory, Washington DC 20375-5000}

\date{\today}

\maketitle

\begin{abstract}
We present a careful recasting of first-principles band structure
calculations for MgB$_2$ in a non-orthogonal sp-tight-binding (TB)
basis.  Our TB results almost exactly reproduce our full potential
linearized augmented plane wave results for the energy bands, the
densities of states and the total energies.  Our procedure generates
transferable Slater-Koster parameters which should be useful for
other studies of this important material.
\end{abstract}
]

The recent discovery of superconductivity in
MgB$_2$\cite{nagamatsu01:_super_k} has created great interest in the
study of this material, both to understand the mechanism of
superconductivity and to explore other properties of MgB$_2$ and
related materials.  Intensive research has been carried out both by
experimentalists\cite{nagamatsu01:_super_k,canfield01:_super,wang01:_specif_k_teslas_k_mgb}
and
theorists.\cite{an01:_super,kortus01:_super,bohnen01:_phonon_mgb_alb,kong01:_elect,liu01:_beyon_elias_mgb,mehl01:_cub2}
There have been several studies of the electronic structure of
MgB$_2$ including total energy, band structure and phonon spectra
calculations as well as evaluations of the electron-phonon coupling,
which seems to have emerged as the prime candidate for explaining
the superconducting behavior.

In this paper we present a highly accurate tight-binding (TB)
description of the band structure and total energy of MgB$_2$.
While there have been TB interpretations of the electronic structure
of MgB$_2$ in the literature, a realistic recasting of the details
of the first-principles electronic structure calculations is
lacking.  Our approach follows the NRL-TB
methodology,\cite{cohen94:_tight,mehl96:_appli} which is based on
deriving a non-orthogonal TB Hamiltonian by fitting to both the
total energy and energy band results of a first-principles
full-potential Linearized Augmented Plane Wave
(LAPW)\cite{andersen75:_linea,singh86:_accel} calculation using the
Hedin-Lundqvist parametrization of the Local Density Approximation
(LDA).\cite{hedin71:_explic_local_exchan_correl_poten} We first
performed detailed LAPW calculations for MgB$_2$ in its ground state
(AlB$_2$) structure, varying c and a, thus determining the LDA
equilibrium volume.  It was necessary to perform 17 independent LAPW
calculations over a large range of volumes and c/a ratios.  Our LAPW
equilibrium parameters are c~=~6.55~a.u. and a~=~5.75~a.u., as compared
to the experimental values of c~=~6.66~a.u. and a~=~5.83~a.u.  As is
usually the case the LDA underestimates the experimental values,
here by about 1.5\%.

\begin{figure}
\epsfig{file=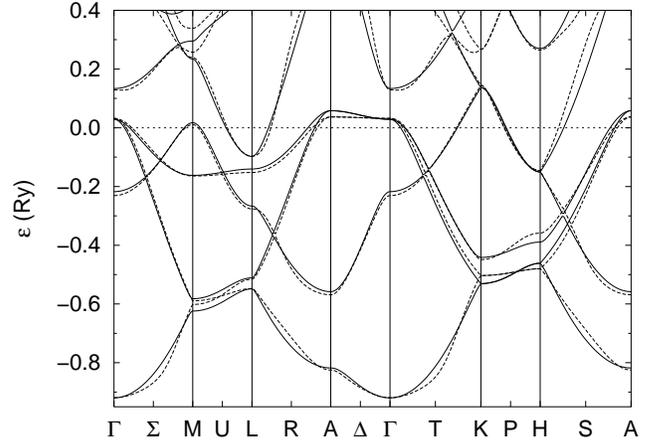,width=3.5in}
\caption{The band structure of MgB$_2$ in the AlB$_2$ structure at
the theoretical equilibrium volume, as determined by the
full-potential LAPW method (solid lines) and our tight-binding
parametrization (dashed lines).  The Fermi level is at $\varepsilon
= 0$.}
\label{fig:mgb2bands}
\end{figure}

All the above results, i.e., 17 values of the total energy and the
energy bands for 76 k-points in the irreducible hexagonal Brillouin
zone, were used as a database to determine the parameters of our
tight-binding Hamiltonian.  According to the NRL-TB scheme the
on-site parameters depend on the density of the neighboring atoms
and the hopping integrals have a polynomial dependence that extends
to at least the third nearest-neighbor distance.  Our basis included
the s and p orbitals on both Mg and B in a non-orthogonal two-center
representation.  A wave-function analysis of our LAPW results shows
that the bands up to the Fermi level, $\varepsilon_F$, are strongly
dominated by the B p states with very little contribution from the
Mg ions.  It turns out, however, that an accurate TB fit including
only the B orbitals is impossible, and therefore the Mg s and p
orbitals were included in the fit.  Furthermore, to obtain a highly
accurate fit it was essential to block diagonalize the Hamiltonian
at the high symmetry points $\Gamma$, A, L, K and H.  We find that
at a given set of lattice parameters (c,a) we can perfectly
reproduce the energy bands of MgB$_2$.  A comparison is shown in
Fig.~\ref{fig:mgb2bands}, where the solid and broken lines represent
the LAPW and TB bands, respectively, at the LDA values of the
equilibrium lattice parameters.  The TB bands are in excellent
agreement with the LAPW bands, including the two-dimensional
B-$\sigma$ band in the $\Gamma \rightarrow A$ direction just above
$\varepsilon_F$, which has been identified as hole band controlling
superconductivity.\cite{kortus01:_super,kong01:_elect} The RMS
fitting error is 2~mRy for the total energy, and close to 10~mRy for
the first five bands.  Beyond the fifth band our fit is not as
accurate, as the Mg $d$-bands, which are not included in our
Hamiltonian, come into play.  The values of our TB parameters are
given in Table~\ref{tab:mgbpar} following the notation of Bernstein
{\em et al.}\cite{bernstein00:_energ} In this table we also show,
for the convenience of the reader, the actual Slater-Koster
parameters for three or four nearest neighbors determined from our
formulas for the specific LDA equilibrium values of the lattice
constants.

\begin{figure}
\epsfig{file=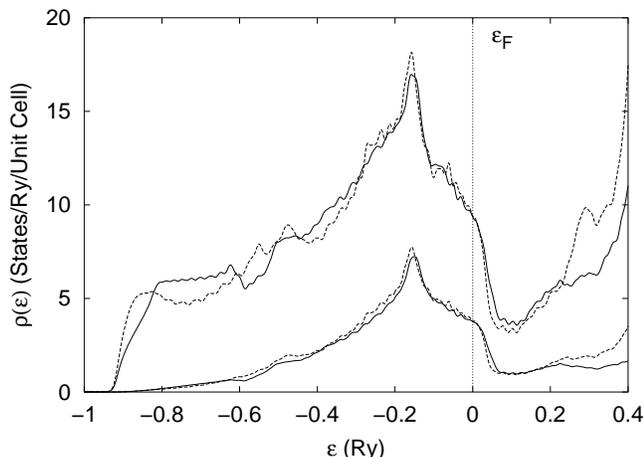,width=3.5in}
\caption{The electronic density of states (DOS) of MgB$_2$ in the
AlB$_2$ structure at the theoretical equilibrium volume, comparing
the total DOS as determined by the full-potential LAPW method (upper
solid line) and our tight-binding parametrization (upper dashed
line), and the partial single-atom B-$p$ decomposition (lower
lines).  The LAPW result decomposition was determined inside the
muffin-tin and then scaled by a factor of 2.37 (see text).  }
\label{fig:mgb2dos}
\end{figure}

In Fig.~\ref{fig:mgb2dos} we show a comparison of TB and LAPW
densities of states.  There is an excellent agreement in both the
total DOS and the B $p$-like DOS.  To facilitate the comparison we
have normalized the muffin-tin decomposed LAPW values so that the
contributions from the angular momentum components add up to the
total DOS, as is the case in the TB.  For the Boron states this
amounted to multiplying the decomposed values by 2.37.  The B
$s$-components of the DOS have their strongest presence at the
bottom of the valence band, from -0.8~Ry to -0.6~Ry on our scale.
They are much smaller than the $p$-like DOS, so we chose not to
include them in Fig.~\ref{fig:mgb2dos}.  Additionally, we have
omitted the Mg $p$-like DOS, which is also small below
$\varepsilon_F$, although it becomes significant above
$\varepsilon_F$.  Our TB value of the total DOS at $\varepsilon_F$
is $N(\varepsilon_F)$~=~0.69~states/eV, which is almost identical to
that found from our direct LAPW calculation.  This value of
$N(\varepsilon_F)$ corresponds to the LDA equilibrium volume and is
slightly smaller than the value of 0.71~states/eV reported by other
workers\cite{an01:_super,kortus01:_super,bohnen01:_phonon_mgb_alb,kong01:_elect,liu01:_beyon_elias_mgb}
at the experimental volume.  Using our value of $N(\varepsilon_F)$
and the measured value\cite{wang01:_specif_k_teslas_k_mgb} of the
specific heat coefficient $\gamma$ we infer a value of
$\lambda$~=~0.65 which is consistent with the high superconducting
transition temperature in MgB$_2$.  It should also be noted that the
B-$p$ states contribute 81\% of the DOS at $\varepsilon_F$.

Our TB-Hamiltonian also provides an accurate description of the
energetics of MgB$_2$, which is expected to be very useful for other
theoretical studies.  We have further tested our parameters by
computing the TB equilibrium structure.  We find an equilibrium of
c~=~6.66~a.u. and a~=~5.79~a.u., in good agreement with the LAPW
result.  At c/a~=~1.14 (the experimental value), we deduce a bulk
modulus of B~=~165~GPa which in good agreement with the experimental
value of 120GPa and with the calculated value of 147~GPa reported by
Bohnen {\em et al.}\cite{bohnen01:_phonon_mgb_alb}

The TB parameters presented in this paper give a very accurate
description of the band structure of MgB$_2$.  The availability of
this Hamiltonian should motivate the calculation of other properties
of this important material.

We thank I. I. Mazin and D. J. Singh for useful discussions.  This
work was supported by the U. S. Office of Naval Research.  The
development of the tight-binding codes was supported in part by the
U. S. Department of Defense Common HPC Software Support Initiative
(CHSSI).


\bibliographystyle{prsty}
\bibliography{mypub2,superconductivity,dft}

\begin{thebibliography}{10}

\bibitem{nagamatsu01:_super_k}
J. Nagamatsu {\it et~al.}, Nature {\bf 410},  63  (2001).

\bibitem{canfield01:_super}
P.~C. Canfield {\it et~al.}, Phys. Rev. Lett. {\bf 86},  2423  (2001).

\bibitem{wang01:_specif_k_teslas_k_mgb}
Y. Wang, T. Plackowski, and A. Junod, Physica C {\bf 355},  179  (2001).

\bibitem{an01:_super}
J.~M. An and W.~E. Pickett, Phys. Rev. Lett. {\bf 86},  4366  (2001).

\bibitem{kortus01:_super}
J. Kortus {\it et~al.}, Phys. Rev. Lett. {\bf 86},  4656  (2001).

\bibitem{bohnen01:_phonon_mgb_alb}
K.-P. Bohnen, R. Heid, and B. Renker, Phys. Rev. Lett. {\bf 86},  5771
  (2001).

\bibitem{kong01:_elect}
Y. Kong, O.~V. Dolgov, O. Jepsen, and O.~K. Andersen, Phys. Rev. B {\bf
  64},  020501  (2001).

\bibitem{liu01:_beyon_elias_mgb}
A.~Y. Liu, I.~I. Mazin, and J. Kortus, condmat/0103570  (2001).

\bibitem{mehl01:_cub2}
M. Mehl, D. Papaconstantopoulos, and D. Singh, cond-mat/0104548  (2001).

\bibitem{cohen94:_tight}
R.~E. Cohen, M.~J. Mehl, and D.~A. Papaconstantopoulos, Phys. Rev. B {\bf
  50},  14694  (1994).

\bibitem{mehl96:_appli}
M.~J. Mehl and D.~A. Papaconstantopoulos, Phys. Rev. B {\bf 54},  4519
  (1996).

\bibitem{andersen75:_linea}
O.~K. Andersen, Phys. Rev. B {\bf 12},  3060  (1975).

\bibitem{singh86:_accel}
D. Singh, H. Krakauer, and C.~S. Wang, Phys. Rev. B {\bf 34},  8391
  (1986).

\bibitem{hedin71:_explic_local_exchan_correl_poten}
L. Hedin and B.~L. Lundqvist, Journal of Physics C {\bf 4},  2064  (1971).

\bibitem{bernstein00:_energ}
N. Bernstein {\it et~al.}, Phys. Rev. B {\bf 62},  4477  (2000).

\end{thebibliography}


\newpage

\widetext

\begin{table}
\caption{Tight-binding parameters for MgB$_2$, generated following
the methods of Mehl and
Papaconstantopoulos\protect\cite{mehl96:_appli} and Bernstein {\em
et al.}\protect\cite{bernstein00:_energ}.  Also shown are the
generated Slater-Koster tight-binding parameters for the nearest
neighbors at the LDA equilibrium lattice constants,
a~=~5.75~a.u. and c~=~6.53~a.u.  On-site energies are generated from
the ``densities'' of like atoms, that is, the Mg on-site parameters
come from the density of Mg atoms, and the B on-site parameters from
the density of B atoms.  $F(R)$ is the cutoff function from equation
(2) of Berstein {\em et al.}\protect\cite{bernstein00:_energ}, with
$R_c$~=~12.5~a.u. and $L_c$~=~0.5a.u.  All energies are in Rydbergs,
all distances in a.u.}
\begin{tabular}{c|rrrr|rrrr}
\multicolumn{9}{c}{Mg-Mg Interactions} \\
\tableline
\multicolumn{9}{c}{On-site Parameters ($\lambda =
0.93961~a.u.^{-1/2}$)} \\
\tableline
$\ell$ & \multicolumn{1}{c}{$\alpha_\ell$} &
\multicolumn{1}{c}{$\beta_\ell$} & \multicolumn{1}{c}{$\gamma_\ell$}
& \multicolumn{1}{c|}{$\chi_\ell$} & \multicolumn{4}{c}{LDA
Equilibrium Values} \\
\tableline
$s$ & 0.02169 & -0.25368 & -0.04017 & 19.84215 &
\multicolumn{4}{c}{0.03516} \\
$p$ & 0.39868 & -0.22303 & 1.35834 & 53.36624 &
\multicolumn{4}{c}{0.52322} \\
\tableline
\multicolumn{9}{c}{Hopping Terms:  $H_{\ell\ell'\mu} (R) =
(a_{\ell\ell'\mu} + b_{\ell\ell'\mu} R + c_{\ell\ell'\mu} R^2) \exp
(-g_{\ell\ell'\mu}^2 R) F(R)$} \\
\tableline
$H_{\ell\ell'\mu}$ &
\multicolumn{1}{c}{$a_{\ell\ell'\mu}$} &
\multicolumn{1}{c}{$b_{\ell\ell'\mu}$} &
\multicolumn{1}{c}{$c_{\ell\ell'\mu}$} &
\multicolumn{1}{c|}{$g_{\ell\ell'\mu}$} & \multicolumn{1}{c}{$a$} &
\multicolumn{1}{c}{$c$} & \multicolumn{1}{c}{$\sqrt3 a$} \\
\tableline
$H(ss\sigma)$ & 5715.097 & -310.8836 & -182.0526 &
1.35579 & -0.05372 & -0.02495 & -0.00009 \\
$H(sp\sigma)$ & 5704288.  & 541286.7 & -387450.5 &
1.84506 & -0.01259 & -0.00161 & -0.00000 \\
$H(pp\sigma)$ & -1920.935 &  498.3775 & -22.58129 &
1.12482 & 0.13720 & 0.09557 & 0.00141 \\
$H(pp\pi)$    &  2000.513 & -739.8181 & 70.26517 & 1.13170 &
0.04414 & 0.03861 & 0.00241 \\
\tableline
\multicolumn{9}{c}{Overlap Terms: $S_{\ell\ell'\mu} (R) =
(\delta_{\ell\ell'} + t_{\ell\ell'\mu} R + q_{\ell\ell'\mu} R^2 +
r_{\ell\ell'\mu} R^3) \exp (-u_{\ell\ell'\mu}^2 R) F(R)$} \\
\tableline
$S_{\ell\ell'\mu}$ &
\multicolumn{1}{c}{$t_{\ell\ell'\mu}$} &
\multicolumn{1}{c}{$q_{\ell\ell'\mu}$} &
\multicolumn{1}{c}{$r_{\ell\ell'\mu}$} &
\multicolumn{1}{c|}{$u_{\ell\ell'\mu}$} & \multicolumn{1}{c}{$a$} &
\multicolumn{1}{c}{$c$} & \multicolumn{1}{c}{$\sqrt3 a$} \\
\tableline
$S(ss\sigma)$ & 1.04886 & -1.27181 & 0.55382 & 1.01629 &
0.18512 & 0.12683 & 0.00767 \\
$S(sp\sigma)$ & 0.41781 & 0.03630 & -0.00873 & 0.63396 &
0.19274 & 0.13373 & -0.00815 \\
$S(pp\sigma)$ & -24.36368 & 0.17541 & 0.41661 & 1.07340 &
-0.07174 & -0.01867 & 0.00101 \\
$S(pp\pi)$ & -68.95974 & 5.97517 & 2.55826 & 1.19338 &
0.08007 & 0.04731 &  0.00088 \\
\tableline\tableline
\multicolumn{9}{c}{B-B Interactions} \\
\tableline
\multicolumn{9}{c}{On-site Parameters ($\lambda =
0.79205~a.u.^{-1/2}$)} \\
\tableline
$\ell$ & \multicolumn{1}{c}{$\alpha_\ell$} &
\multicolumn{1}{c}{$\beta_\ell$} & \multicolumn{1}{c}{$\gamma_\ell$}
& \multicolumn{1}{c|}{$\chi_\ell$} & \multicolumn{4}{c}{LDA
Equilibrium Values} \\
\tableline
$s$ & -0.16521 & -0.00022 & 0.02579 & 0.09088
& \multicolumn{4}{c}{-0.09356} \\
$p$ & 0.38802 & 0.00060 & 0.00566 & 0.01918 &
\multicolumn{4}{c}{0.40383} \\
\tableline
\multicolumn{9}{c}{Hopping Terms:  $H_{\ell\ell'\mu} (R) =
(a_{\ell\ell'\mu} + b_{\ell\ell'\mu} R + c_{\ell\ell'\mu} R^2) \exp
(-g_{\ell\ell'\mu}^2 R) F(R)$} \\
\tableline
$H_{\ell\ell'\mu}$ &
\multicolumn{1}{c}{$a_{\ell\ell'\mu}$} &
\multicolumn{1}{c}{$b_{\ell\ell'\mu}$} &
\multicolumn{1}{c}{$c_{\ell\ell'\mu}$} &
\multicolumn{1}{c|}{$g_{\ell\ell'\mu}$} &
\multicolumn{1}{c}{$a/\sqrt3$} & \multicolumn{1}{c}{$a$} &
\multicolumn{1}{c}{$c$} & \multicolumn{1}{c}{$2a/\sqrt3$} \\
\tableline
$H(ss\sigma)$ & -7.31550 & 2.09241 & -0.23379 & 0.85573 &
-0.25908 & -0.04471 & -0.03032 & -0.02881\\
$H(sp\sigma)$ & -146.6733 & 64.75572 & -8.57386 & 1.21983 &
-0.18743 & -0.01112 & -0.00538 & - 0.00484 \\
$H(pp\sigma)$ & -296.2214 & 128.0942 & -10.42990 & 1.17220
& 0.14703 & 0.03537 & 0.01210 & 0.01030 \\
$H(pp\pi)$ & 167.1287 & -84.95581 & 9.40729 & 1.16057 &
-0.12834 & -0.00448 & 0.00204 & 0.00232 \\
\tableline
\multicolumn{9}{c}{Overlap Terms: $S_{\ell\ell'\mu} (R) =
(\delta_{\ell\ell'} + t_{\ell\ell'\mu} R + q_{\ell\ell'\mu} R^2 +
r_{\ell\ell'\mu} R^3) \exp (-u_{\ell\ell'\mu}^2 R) F(R)$} \\
\tableline
$S_{\ell\ell'\mu}$ &
\multicolumn{1}{c}{$t_{\ell\ell'\mu}$} &
\multicolumn{1}{c}{$q_{\ell\ell'\mu}$} &
\multicolumn{1}{c}{$r_{\ell\ell'\mu}$} &
\multicolumn{1}{c|}{$u_{\ell\ell'\mu}$} &
\multicolumn{1}{c}{$a/\sqrt3$} & \multicolumn{1}{c}{$a$} &
\multicolumn{1}{c}{$c$} & \multicolumn{1}{c}{2$a/\sqrt3$} \\
\tableline
$S(ss\sigma)$ & 0.08974 & -0.05865 & 0.00446 & 0.60130 &
0.24535 & 0.05320 & 0.03092 & 0.02870 \\
$S(sp\sigma)$ & 14.02893 & -2.43293 & 0.33914 & 1.27302 &
0.14823 & 0.00581 & 0.00208 & 0.00181 \\
$S(pp\sigma)$ & -60.70629 & -0.47590 & 3.29498 & 1.32165 &
-0.25833 & 0.01141 & 0.00558 & 0.00497 \\
$S(pp\pi)$ & 18.98764 & 6.13369 & -3.44742 & 1.41889 &
0.00689 & -0.00321 & -0.00112 & -0.00096 \\
\tableline\tableline
\multicolumn{9}{c}{Mg-B Interactions} \\
\tableline
\multicolumn{9}{c}{Hopping Terms:  $H_{\ell\ell'\mu} (R) =
(a_{\ell\ell'\mu} + b_{\ell\ell'\mu} R + c_{\ell\ell'\mu} R^2) \exp
(-g_{\ell\ell'\mu}^2 R) F(R)$} \\
\tableline
$H_{\ell\ell'\mu}$ &
\multicolumn{1}{c}{$a_{\ell\ell'\mu}$} &
\multicolumn{1}{c}{$b_{\ell\ell'\mu}$} &
\multicolumn{1}{c}{$c_{\ell\ell'\mu}$} &
\multicolumn{1}{c|}{$g_{\ell\ell'\mu}$} &
\multicolumn{1}{c}{$\sqrt{\frac13 a^2 +\frac14 c^2}$} & 
\multicolumn{1}{c}{$\sqrt{\frac43 a^2 +\frac14 c^2}$} &
\multicolumn{1}{c}{$\sqrt{\frac73 a^2 +\frac14 c^2}$} \\
\tableline
$H(ss\sigma)$ & -15.40626 & 8.92332 & -2.25890 & 1.06263 &
-0.11887 & -0.01709 & -0.00257 \\
$H(sp\sigma)$ & -22.65145 &  5.35089 & -0.60679 & 1.03205 &
-0.07642 & -0.00612 & -0.00093 \\
$H(pp\sigma)$ & 98.38228 &-45.01479 &  6.05711 & 1.20823 &
0.02245 & 0.00196 & 0.00019 \\
$H(pp\pi)$ & -94.47230 & 33.60639 & -4.25418 & 1.21106 &
-0.03269 &-0.00152 &-0.00013 \\
$H(ps\sigma)$ & 7.80580 &  1.71300 & -0.22442 & 1.03201 &
0.07662 & 0.00308 & 0.00015 \\
\tableline
\multicolumn{9}{c}{Overlap Terms: $S_{\ell\ell'\mu} (R) =
(t_{\ell\ell'\mu} + q_{\ell\ell'\mu} R + r_{\ell\ell'\mu} R^2) \exp
(-u_{\ell\ell'\mu}^2 R) F(R)$} \\
\tableline
$S_{\ell\ell'\mu}$ &
\multicolumn{1}{c}{$t_{\ell\ell'\mu}$} &
\multicolumn{1}{c}{$q_{\ell\ell'\mu}$} &
\multicolumn{1}{c}{$r_{\ell\ell'\mu}$} &
\multicolumn{1}{c|}{$u_{\ell\ell'\mu}$} &
\multicolumn{1}{c}{$\sqrt{\frac13 a^2 + \frac14 c^2}$} & 
\multicolumn{1}{c}{$\sqrt{\frac43 a^2 + \frac14 c^2}$} &
\multicolumn{1}{c}{$\sqrt{\frac73 a^2 + \frac14 c^2}$} \\
\tableline
$S(ss\sigma)$ & 1.74820 & 0.13546 & 0.07434 & 0.82425 &
0.16873 & 0.04450 & 0.01278 \\
$S(sp\sigma)$ &	15.27243 & 3.72217 & 0.27051 & 1.09173 &
0.14959 & 0.00848 & 0.00081 \\
$S(pp\sigma)$ &	-4.51769 & -4.96427 & 0.71842 & 0.95339 &
-0.17503 & -0.00229 & 0.00188 \\
$S(pp\pi)$ & 846.58108 & 265.43883 & -10.03535 & 1.30600 &
0.06108 & 0.00187 & 0.00007 \\
$S(ps\sigma)$ & -2.81156 & 0.20700 & -0.40404 & 0.92459 &
-0.19812 & -0.04168 & -0.00940 \\
\end{tabular}
\label{tab:mgbpar}
\end{table}

\end{document}